# Distinct transport behaviors of $LaFe_{1-y}Co_yAsO_{1-x}F_x$ ($x$=0.11) between the superconducting and nonsuperconducting metallic $y$ regions divided by $y$ ~ 0.05

S. C. Lee,[1] A. Kawabata,[1] T. Moyoshi,[1] Y. Kobayashi[1,2] and M. Sato[1,2,] *

[1]*Department of Physics, Division of Material Science, Nagoya University, Furo-cho, Chikusa-ku, Nagoya 464-8602*

[2]*JST, TRIP, Nagoya University, Furo-cho, Chikusa-ku, Nagoya 464-8602*

Electrical resistivities, Hall coefficients and thermoelectric powers have been measured for polycrystalline samples of $LaFe_{1-y}Co_yAsO_{1-x}F_x$ ($x$=0.11) with various values of $y$. The results show that there exists clear distinction of these transport behaviors between the superconducting and nonsuperconducting metallic regions of $y$ divided by the boundary value $y_c$~0.05. We have found that the behaviors in both regions are very similar to those of high-$T_c$ Cu oxides in the corresponding phases. If they reflect, as in the case of Cu oxides, effects of strong magnetic fluctuations, the energy scale of the fluctuations is considered to be smaller than that of the high Cu oxides by a factor of ~1/2. Arguments on the electronic nature and superconducting symmetry are presented on the basis of the observed small rate of the $T_c$ suppression rate by the Co doping.

Superconductivity found in $LaFeAsO_{1-x}F_x$ with FeAs conducting layers formed of corner-sharing $FeAs_4$ tetrahedra [1)] has renewed questions to what extent of materials and to what high temperature, we can find superconductivity, and by search works for new superconductors with similar structures, $T_c$ values higher than 50 K have been found for systems derived by total substitutions of La with various other lanthanide elements Ln, (1111 system).[2)] Systems derived from $AFe_2As_2$ (A= alkali earth elements, 122 system) have also been found to be superconducting.[3)] In addition, FeSe, which has layers formed of corner-sharing $FeSe_4$ tetrahedra, was found to be superconducting.[4, 5)] Because they have layers FeAs or FeSe layers, as the stage of the superconductivity, strong correlation of 3$d$ electrons in these layers seem, at first sight, to be important in arguing the origin of the rather high $T_c$ values. The occurrence of the spin-density-wave (SDW)-like transition in the mother system LaFeAsO [6)] also suggests that the magnetic interaction of these systems cannot be ignored in the arguments.

To investigate the electronic state of FeAs layers, we prepared $LaFe_{1-y}Co_yAsO_{1-x}F_x$ and carried out various kinds of studies by measuring transport and magnetic properties, NMR Knight shifts and nuclear relaxation rates and so on. Several results of these studies were reported previously,[7-9)] where we first pointed out that Co atoms doped to Fe sites do not act as the pair breaking centers, indicating that the superconducting order parameter is nodeless. We also pointed out that even for the order parameters with opposite signs on disconnected Fermi surfaces, the observed insensitiveness of $T_c$ to the nonmagnetic Co impurities was hard to simply explain. The appearance of the superconducting transitions reported after ref. 7 in samples of $LaFe_{1-y}Co_yAsO$ and $AFe_{2-y}Co_yAs_2$ (A=Ba, Sr) with $y$ as large as 0.1-0.2[10-13)] supports the idea stated above. We have also shown that in the superconducting phase the singlet pairing is realized.[9)]

In the present letter, we mainly report results of the transport studies and show that there exists clear distinction of the transport behaviors between the superconducting and nonsuperconducting metallic regions of $y$ divided by the boundary value $y_c$~0.05. This characteristic feature of the transport behaviors is very similar to that of high-$T_c$ Cu oxides.[14, 15)]

Polycrystalline samples of $LaFe_{1-y}Co_yAsO_{1-x}F_x$ ($x$ always being fixed at 0.11) were prepared from initial mixtures of La, $La_2O_3$, $LaF_3$ and FeAs with the nominal molar ratios. Details of the preparation processes can be found in the previous papers.[8, 9)] The X-ray powder patterns were taken with Cukα radiation. The superconducting diamagnetic moments were measured by a Quantum Design SQUID magnetometer with the magnetic field $H$ of 10 G under both conditions of the zero-field-cooling (ZFC) and field cooling (FC). From the data of the electrical resistivities ρ and diamagnetic moments, the $T_c$ values were determined as described poreviously,[8, 9)] where we found that both kinds of $T_c$ values agree well.

Hall coefficients $R_H$ of the polycrystalline samples were measured with increasing $T$ stepwise under the magnetic field $H$ of 7 T, where the sample plates were rotated around an axis perpendicular to the field, and thermoelectric powers $S$ were measured by the methods described in refs. 14 and 15.

The top panel of Fig. 1 shows the examples of the X-ray powder patterns taken with Cukα radiation for the samples with $y$=0.0 and 0.3, where the calculated pattern of LaFeAsO is also shown. Very weak peaks of impurity phases were found at the scattering angle 2θ of 26.69° and 27.82° and they were identified as the contribution from LaOF and LaAs, respectively. The molar fractions of these impurity phases were estimated to be ~3 % for the former and ~2.6 % for the latter. Any other impurity phases have not been detected. The linear *y*-dependence of the lattice parameter *c* shown in the bottom left panel of Fig. 1 guarantees that the Co doping into the Fe sites was successfully carried out.[7-9] The data of the superconducting diamagnetism taken for samples of $LaFe_{1-y}Co_yAsO_{1-x}F_x$ under the ZFC condition are shown in the bottom right panel of Fig. 1, where the 100*y* values are attached to the corresponding data. (Note that the $T_c$ values do not correlate with *y*, even though we know that the Co doping was successfully carried out. It indicates that the doped Co atoms do not act as the pair breaking centers. See the previous papers[8, 9] for details.)

*Corresponding author: e43247a@nucc.cc.nagoya-u.ac.jp

Figure 2 shows the *T* dependence of the electrical resistivities of the samples with various *y* values. The superconducting transition appears in the region of $y<y_c\sim$ 0.05, and samples with $y\geq$ 0.1 exhibit nonsuperconducting metallic behavior. An interesting point is that the ρ-*T* curves can be categorized into two groups, that is, the room-temperature resistivities ρ observed for $y\geq$ 0.1 have significantly smaller values, indicating that the electronic states of these two regions seems to be different.

The above idea is supported by the data of the Hall coefficient $R_H$ and thermoelectric power *S* of these samples, which are shown in Figs. 3 and 4. In Fig. 3, we can see that the samples with $y\leq 0.05$ exhibits rather strong *T* dependences of $R_H$, while for the samples with $y>y_c$, the absolute values of $R_H$ are small and have weak *T* dependence. These behaviors of $R_H$ are very similar to those of high-$T_c$ Cu oxides,[14] in which effects of active magnetism or strong magnetic fluctuation plays an important role in the superconducting region of *y*.[16] The data shown in Fig. 4 for the thermoelectric power *S* also have features similar to those of high-$T_c$ Cu oxides,[15] that is, for $y<y_c$, the *S*-*T* curves exhibit the characteristic structure similar to that observed for high-$T_c$ Cu oxides, while $y>y_c$, they exhibit, roughly speaking, ordinary *T*-linear behavior. These similarities suggest that in the region $y\leq 0.05$, the present system has rather unusual electronic state realized by the active magnetism. It is, of course, consistent with the facts that an antiferromagnetic (or SDW) transition exists in the system of LaFeAsO and that for $LaFe_{1-y}Co_yAsO_{1-x}F_x$ ($x$=0.11), significant effects of the antiferromagnetic fluctuation can be observed in the $^{75}$As NMR spectra and transverse relaxation curves of the samples with $y<y_c$.[8,9] The energy or temperature scale of the magnetic fluctuations observed here seems to be smaller than that of high $T_c$ Cu oxides by a factor of ~1/2. If the magnetic fluctuations are important for the occurrence of the superconductivity, this factor gives us the possible maximum $T_c$ value of systems with FeAs layers.

We have so far presented the results of the transport measurements, where effects of Co-doping to Fe sites of $LaFeAsO_{1-x}F_x$ ($x$=0.11) have been mainly focused on. As the result of the first stage of the study, we have found that Co atoms do not carry localized moments in $LaFe_{1-y}Co_yAsO_{1-x}F_x$. This result can also be extracted from the fact that we have not observed the significant $T_c$ suppression expected for systems suffering the electron scattering by localized magnetic moments. (For superconductors with nodes and even for superconductors without nodes but with sign difference of the order parameters among disconnected Fermi surfaces, the potential scattering due to impurities is ,naively speaking, suppresses $T_c$ significantly, too.) The absence of magnetic moment at Co site has turned out to be consistent with the result of NMR studies by Ning *et al.*[17]

Based on the absence of localized moments at Co sites, we may be able to adopt a rigid band picture, and if we forget, for a while, the effects of the potential scattering due to Co impurities, the doping can be considered just to change the carrier numbers of the system, as in the case of F-substitution for oxygen atoms. In this case, the carrier number of $LaFe_{1-y}Co_yAsO_{1-x}F_x$ corresponds to that of $LaFeAsO_{1-x-y}F_{x+y}$. For $LaFe_{1-y}Co_yAsO_{1-x}F_x$, we have found that the superconductivity appears for $(x+y)\leq$ 0.16, which is consistent with the phase diagram reported by Kamihara *et al.* for $LaFeAsO_{1-x}F_x$. According to the band calculation,[18, 19] on the way from LaFeAsO to LaCoAsO or from $(x+y)$=0 to 1.0, the electronic density of states $N(\varepsilon_F)$ at the chemical potential μ first decreases steeply to a very small value, then gradually increases and passes a sharp peak. We think that the first decrease of $N(\varepsilon_F)$ with increasing μ corresponds to the process of burying the hole Fermi surface around the Γ point with large effective mass. In this region, we speculate that the superconductivity appears. With further increase of μ or $(x+y)$, only the Fermi surface around the M point may remain. In this state, the resistivity is small as compared with the superconducting region of μ or $(x+y)$, because the effective mass of the electrons is small and the inter-Fermi-surface-scattering of the electrons do not exist. If we stand on this idea, the occurrence of the superconductivity may be considered to be related to this scattering.[20] Anomalous *T* dependence of $R_H$ and *S* observed for $(x+y)\leq$ 0.16 can also be understood by this idea in a naturally way, because the active magnetism observed in the NMR studies[7-9] is considered to be due to the hole Fermi surface around the Γ point with large effective mass, and because this active magnetism induces the anomalous behaviors of $R_H$ and *S*, similarly to the case of high $T_c$ Cu oxides.

In the above arguments, we have assumed that the potential scattering does not suppress the superconducting transition. To see if this assumption is appropriate or not, the symmetry of the order parameter becomes very important. If the symmetry is *s*-like, the Anderson theorem[21)] is valid and $T_c$ is not suppressed. It is consistent with the appearance of the superconductivity in $Ba(Fe_{1-x}Co_x)_2As_2$ with $x$ as large as 0.14.[22-24)]

Many groups suggested the possibility of the so-called $s_\pm$ symmetry of the order parameter, considering the spin-fluctuation mechanism of the superconductivity.[25)] However, the situation may not be so simple. Because the $T_c$ suppression by impurity scattering cannot be, naively speaking, ignored for the $s_\pm$ symmetry, the observed insensitiveness of $T_c$ to the Co doping seems to contradict the symmetry. Although Senga and Kontani[26)] has argued, considering details of the inter band scattering, possible smallness of $T_c$ suppression rate, the theory has to explain, we think, other kinds of experimental data, such as the $T$ dependence of the NMR longitudinal relaxation rate $1/T_1$ simultaneously with the suppression rate,[27-29)] which seems to remain as an issue to be solved.

Neutron inelastic measurements seem to be able to present information to distinguish which one of *s*- or $s_\pm$ symmetry is realized in the present system. Actually, Christianson *et al.*[30)] have carried out such experiments for a powder sample of $Ba_{1-x}K_xFe_2As_2$ ($x$=0.4) by using the pulse neutron source and suggested the existence of the resonance peak. Such works are underway in our group, too, by using a triple axis neutron spectrometer to obtain a rigid data.

Finally, we make a brief comment on the observation of small spontaneous magnetic moments for samples with $y \geq 0.05$. The values of the spontaneous moments were estimated to be 0.02, 0.07, 0.19 and 0.19 $\mu_B/(Fe_{1-y}Co_y)$ and Curie temperatures are about 40, 105, 360 and310 K for $y$= 0.05, 0.10, 0.25 and 0.40, respectively. As shown in the top panel of Fig. 1, the impurity phases detected by the X-ray measurements are only LaOF and LaAs. The observed Curie temperatures are clearly $y$ dependent, which cannot be explained by considering certain impurity phases, which contain Fe and/or Co atoms. Moreover, the observed ferromagnetic behaviors exhibit characteristic features of weak itinerant ferromagnets, though its description is out of the scope of the present paper. Therefore, it is very difficult to attribute the spontaneous magnetization to the existence of impurity phases. Even if the observed spontaneous magnetization is due to the existence of impurity phase, the essential features of the transport behavior shown above do not change, because the dominant part of the samples is $LaFe_{1-y}Co_yAsO_{1-x}F_x$.

In summary, we have presented the results of the transport measurements of $LaFe_{1-y}Co_yAsO_{1-x}F_x$, which shows the distinct transport behaviors of $LaFe_{1-y}Co_yAsO_{1-x}F_x$ ($x$=0.11) between the superconducting and nonsuperconducting metallic $y$ regions divided by $y \sim 0.05$.

Acknowledgments –The work is supported by Grants-in-Aid for Scientific Research from the Japan Society for the Promotion of Science (JSPS), Grants-in-Aid on Priority Area from the Ministry of Education, Culture, Sports, Science and Technology and JST, TRIP.

References

1) Y. Kamihara, T. Watanabe, M. Hirano, and H. Hosono; J. Am. Chem. Soc. **130** (2008) 3296.
2) Zhi-An Ren, Jie Yang, Wei Lu, Wei Yi, Xiao-Li Shen, Zheng-Cai Li, Guang-Can Che, Xiao-Li Dong, Li-Ling Sun, Fang Zhou, and Zhong-Xian Zhao; Europhys. Lett. **82** (2008) 57002.
3) M. Rotter, M. Tegel, and D. Johrendt, Phys. Rev. Letters **101** (2008) 107006.
4) F-C. Hsu, J-Y. Luo, K-W. Yeh, T-K. Chen, T-W. Huang, P. M. Wu, Y-C. Lee, Y-L. Huang, Y-Y. Chu, D-C. Yan, and M-K. Wu, Proceedings of the National Academy of Sciences of the United States of America **105** (2008) 14262.
5) Y. Mizuguchi, F. Tomioka, S. Tsuda, T. Yamaguchi, Y. Takano, Appl. Phys. Lett. **93** (2008)152505.
6) Clarina de la Cruz, Q. Huang, J. W. Lynn, Jiying Li, W. Ratcliff II, J. L. Zarestky, H. A. Mook, G. F. Chen, J. L. Luo, N. L. Wang, and Pengcheng Dai; Nature **453** (2008) 899.
7) A. Kawabata, S-C. Lee, T. Moyoshi, Y. Kobayashi, and M. Sato; International Symposium on Fe-oxipnictide Superconductors, June 28th-29th, 2008, Kokuyo Hall, Shinagawa, Tokyo, Japan.
8) A. Kawabata, S-C. Lee, T. Moyoshi, Y. Kobayashi, and M. Sato; J. Phys. Soc. Jpn. **77** (2008) 103704.
9) A. Kawabata, S-C. Lee, T. Moyoshi, Y. Kobayashi, and M. Sato; J. Phys. Soc. Jpn. **77** (2008) Supplement C, 147.
10) A. S. Sefat, A. Huq, M. A. McGuire, R. Jin, B. C. Sales, D. Mandrus; Phys. Rev. B **78** (2008) 104505.
11) G. Cao, C. Wang, Z. Zhu, S. Jiang, Y. Luo, S. Chi, Z. Ren, Q.Tao,, Y. Wang, Z. Xu; arXiv: 0807.1304.
12) A. S. Sefat, R. Jin, M. A. McGuire, B. C. Sales, D. J. Singh, and D. Mandrus; Phys. Rev. Lett. **101** (2008) 117004.
13) A. Leithe-Jasper, W. Schnelle, C. Geibel, and H. Rosner; Phys. Rev. Lett. **101** (2008) 207004.
14) T. Nishikawa, J. Takeda, and M. Sato; J. Phys. Soc. Jpn. **63** (1994) 1441.
15) J. Takeda, T. Nishikawa, and M. Sato; Physica C 231 (1994) 293-299.
16) H. Kontani, K. Kanki, and K. Ueda; Phys. Rev. B **59** (1999) 14723.
17) F. Ning, K. Ahilan, T. Imai, A. S. Sefat, R. Jin, M. A. McGuire, B. C. Sales, and D. Mandrus; J. Phys. Soc. Jpn. **77** (2008) 103705.
18) D. J. Singh and M.-H. Du; Phys Rev. Lett. **100** (2008) 237003.
19) H. Li, J. Li, S. Zhang, W. Chu, D. Chen, and Z. Wu; arXiv:0807.3153v2
20) We thank Prof. M. Imada for pointing out this possibility.
21) P. W. Anderson; J. Phys. Chem. Solids 11 (1959) 26.
22) F. Ning, K. Ahilan, T. Imai, Athena S. Sefat, R. Jin, M. A. McGuire, B. C. Sales, and D. Mandrus; arXiv:0811.1617.
23) J-H Chu, J. G. Analytis, C. Kucharczyk, I. R. Fisher; arXiv 0811.2463.
24) X. F. Wang, T. Wu, G. Wu, R. H. Liu, H. Chen, Y. L. Xie, X. H. Chen; arXiv:0811.2920.

25) Kazuhiko Kuroki, Seiichiro Onari, Ryotaro Arita, Hidetomo Usui, Yukio Tanaka, Hiroshi Kontani, and Hideo Aoki; Phys. Rev. Lett. **101** (2008) 087004.
26) Y. Senga and H. Kontani; J. Phys. Soc. Jpn. **77** (2008) 113710.
27) Y. Nakai, K. Ishida, Y. Kamihara, M. Hirano, and H. Hosono; J. Phys. Soc. Jpn. **77** (2008) 073701.
28) H. Mukuda, N. Terasaki, H. Kinouchi, M. Yashima, Y. Kitaoka, S. Suzuki, S. Miyasaka, S. Tajima, K. Miyazawa, P. Shirage, H. Kito, H. Eisaki, and A. Iyo; J. Phys. Soc. Jpn. **77** (2008) 093704.
29) S. Kawasaki, K. Shimada, G. F. Chen, J. L. Luo, N. L. Wang, Guo-qing Zheng; arXiv:0810.1818v2.
30) A. D. Christianson, E. A. Goremychkin, R. Osborn, S. Rosenkranz, M. D. Lumsden, C. D. Malliakas, l. S. Todorov, H. Claus, D. Y. Chung, M. G. Kanatzidis, R. I. Bewley, T. Guidi; arXiv:0807.3932.

## Figure captions

Fig. 1 (top) Examples of the X-ray powder patterns of the samples of $LaFe_{1-y}Co_yAsO_{1-x}F_x$ ($x$=0.11) for $y$=0.0 and 0.3 are shown together with the calculated pattern for $y$=0.0 Weak peaks from impurity phases of LaOF and LaAs are observed at 2θ= 26.69° and 27.82°, respectively. The molar fraction of the impurity phases are estimated to be ~3 % and ~2.6 %**,** respectively. (bottom left) The lattice parameter $c$ is shown against $y$ for $LaFe_{1-y}Co_yAsO_{1-x}F_x$ ($x$=0.11). (bottom right) Magnetic susceptibilities of the powder samples of $LaFe_{1-y}Co_yAsO_{1-x}F_x$ ($x$=0.11) are shown against $T$, where the 100$y$ value is attached to each corresponding sample. They were obtained under the ZFC condition at $H$=10 G.

Fig. 2 Electrical resisitivies of the samples of $LaFe_{1-y}Co_yAsO_{1-x}F_x$ ($x$=0.11) are shown against $T$. To each curve, the 100$y$ value is attached.

Fig. 3 Hall coefficient $R_H$ of the samples of $LaFe_{1-y}Co_yAsO_{1-x}F_x$ ($x$=0.11) are shown against $T$ for various $y$ values. To each curve, the 100$y$ value is attached.

Fig. 4 Thermoelectric powers $S$ of the samples of $LaFe_{1-y}Co_yAsO_{1-x}F_x$ ($x$=0.11) are shown against $T$ for various $y$ values. To each curve, the 100$y$ value is attached.

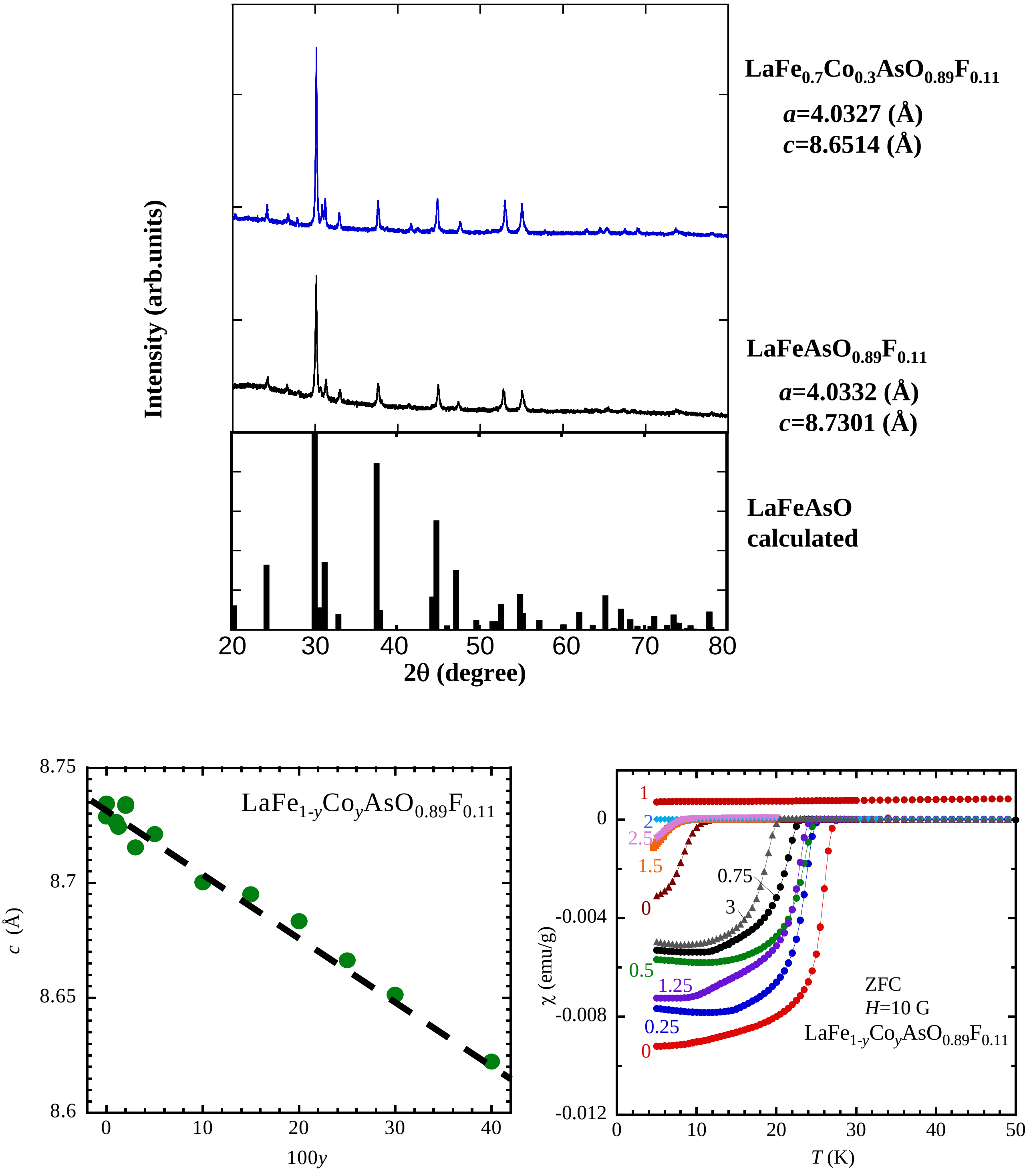

Fig. 1

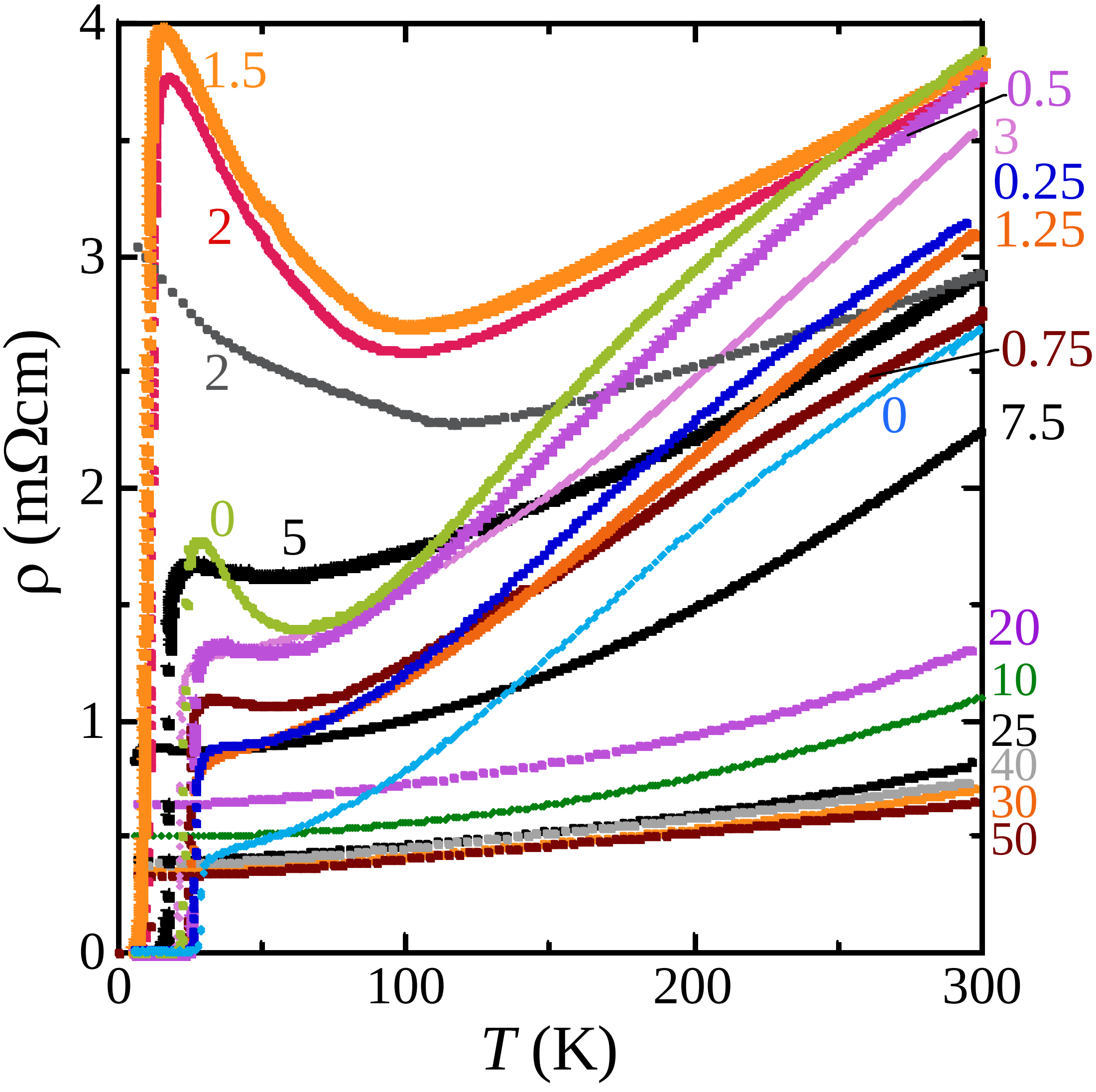

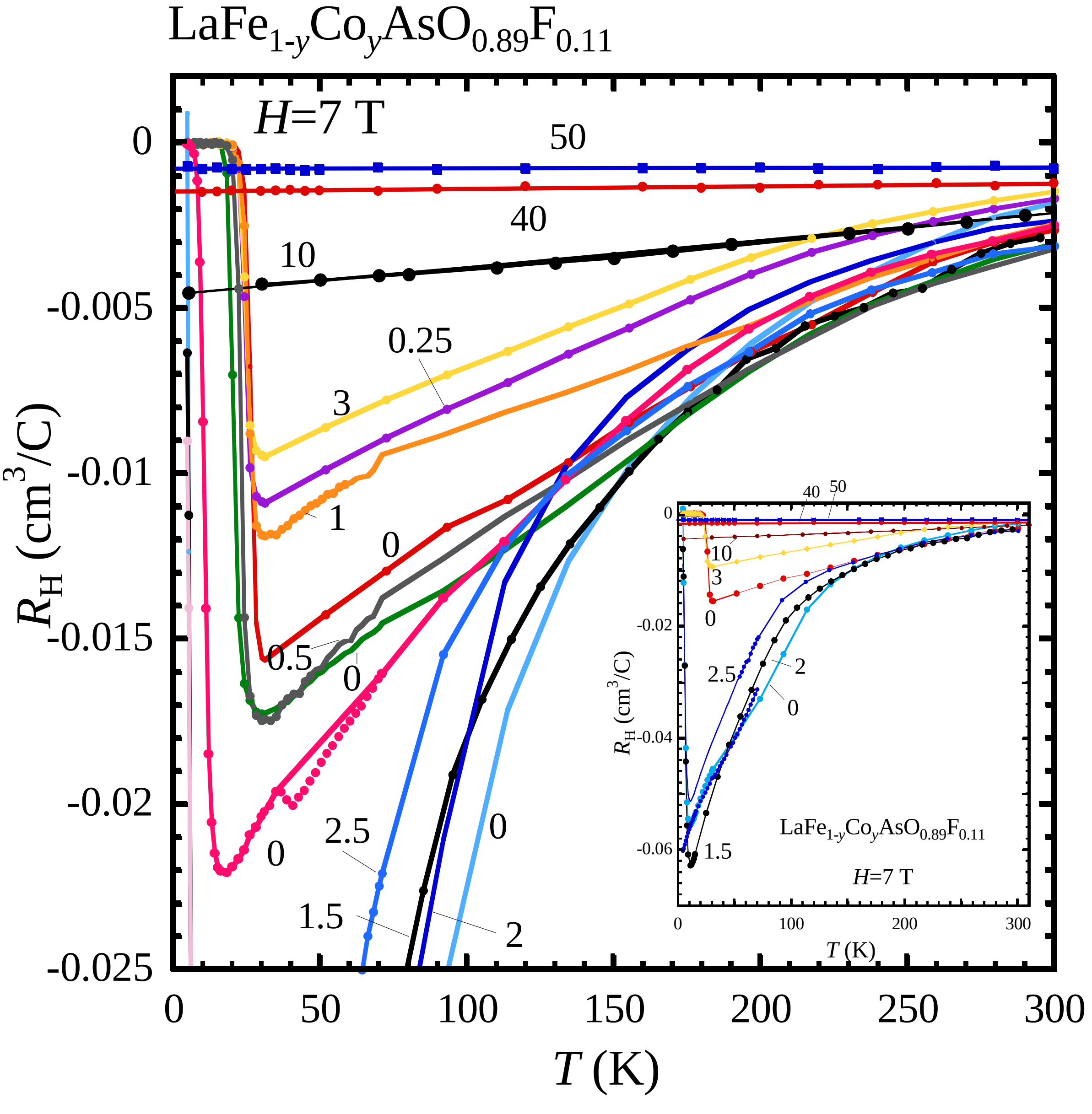

Fig. 3

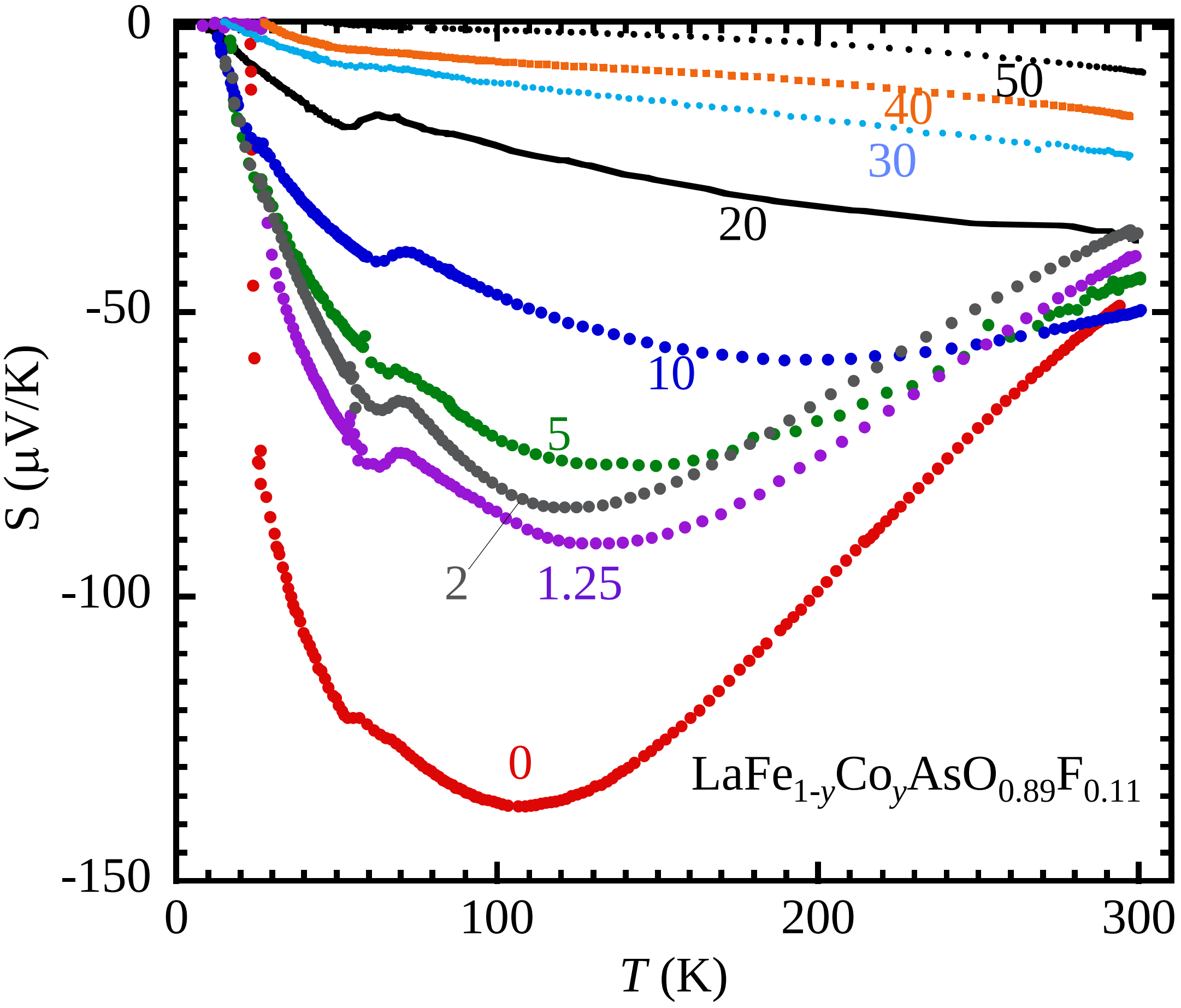

Fig. 4